\documentclass[review]{elsarticle}

\usepackage{hyperref}

\journal{Mechanical Systems and Signal Processing}

\usepackage{amssymb}
\usepackage{graphics}
\usepackage{amsmath,bm}
\usepackage{subfigure}
\usepackage{epstopdf} 
\usepackage{bbold}
\usepackage[bbgreekl]{mathbbol}
\usepackage{color}
\usepackage[compatibility=false]{caption}
\usepackage{setspace}
\usepackage{xcolor}
\definecolor{test}{rgb}{0.7,0.7,1}
\definecolor{test2}{rgb}{1,0.7,0.7}
\usepackage{mathrsfs}
\usepackage{stmaryrd}
\usepackage{mathtools}
\usepackage{placeins}
\usepackage{tikz}
\usepackage{etoolbox}
\usepackage{anyfontsize}

\usepackage{array}
\newcolumntype{L}[1]{>{\centering\let\newline\\\arraybackslash\hspace{0pt}}m{#1}}
\newcolumntype{C}[1]{>{\centering\let\newline\\\arraybackslash\hspace{0pt}}m{#1}}
\newcolumntype{R}[1]{>{\centering\let\newline\\\arraybackslash\hspace{0pt}}m{#1}}

\DeclareSymbolFontAlphabet{\mathbb}{AMSb}
\DeclareSymbolFontAlphabet{\mathbbl}{bbold}

\newrobustcmd*{\mysquare}[1]{\tikz{\filldraw[draw=#1,fill=#1] (0,0) rectangle (0.1cm,0.1cm);}}

\newrobustcmd*{\mycircle}[1]{\tikz{\filldraw[draw=#1,fill=#1] (0,0) circle [radius=0.075cm];}}

\newrobustcmd*{\mytriangle}[1]{\tikz{\filldraw[draw=#1,fill=#1] (0,0) -- (0.15cm,0) -- (0.075cm,0.15cm);}}

\newcommand\BibTeX{{\rmfamily B\kern-.05em \textsc{i\kern-.025em b}\kern-.08em
		T\kern-.1667em\lower.7ex\hbox{E}\kern-.125emX}}

\renewcommand{\SS}{\Theta}

\newcommand{\SSpt}{\theta}

\newcommand{\SA}{\mathbbl{\Sigma}}

\newcommand{\PM}{\mathbb{P}}

\newcommand{\randvar}[1]{\mathbb{#1}}

\newcommand{\stddev}[1]{\sigma_{#1}}

\newcommand{\Z}{\ensuremath{\mathbb{Z}}}           

\begin{document}

\begin{frontmatter}

\title{Damage detection in an uncertain nonlinear beam based on stochastic Volterra series: an experimental application}

\author{Luis G. G. Villani}\ead{luis.villani@unesp.br}
\author{ Samuel da Silva}\ead{samuel.silva13@unesp.br}
\address{UNESP - Universidade Estadual Paulista, Faculdade de Engenharia de Ilha Solteira, Departamento de Engenharia Mec\^anica, Av. Brasil, 56, Ilha Solteira, 15385-000, SP, Brasil}
\author{Americo Cunha Jr.}\ead{americo@ime.uerj.br}
\address{UERJ - Universidade do Estado do Rio de Janeiro, NUMERICO -- Nucleus of Modeling and Experimentation with Computers, R. S\~ao Francisco Xavier, 524, Rio de Janeiro, 20550-900, RJ, Brasil}
\author{Michael D. Todd}\ead{mdtodd@mail.ucsd.edu}
\address{UCSD - University of California San Diego, Department of Structural Engineering, 9500 Gilman Dr, La Jolla, CA, USA}


\cortext[mycorrespondingauthor]{Luis Gustavo Giacon Villani}



\begin{abstract}
The damage detection problem becomes a more difficult task when the intrinsically nonlinear behavior of the structures and the natural data variation are considered in the analysis because both phenomena can be confused with damage if linear and deterministic approaches are implemented. Therefore, this work aims the experimental application of a stochastic version of the Volterra series combined with a novelty detection approach to detect damage in an initially nonlinear system taking into account the measured data variation, caused by the presence of uncertainties. The experimental setup is composed by a cantilever beam operating in a nonlinear regime of motion, even in the healthy condition, induced by the presence of a magnet near to the free extremity. The damage associated with mass changes in a bolted connection (nuts loosed) is detected based on the comparison between linear and nonlinear contributions of the stochastic Volterra kernels in the total response, estimated in the reference and damaged conditions. The experimental measurements were performed on different days to add natural variation to the data measured.  The results obtained through the stochastic proposed approach are compared with those obtained by the deterministic version of the Volterra series, showing the advantage of the stochastic model use when we consider the experimental data variation with the capability to detect the presence of the damage with statistical confidence. Besides, the nonlinear metric used presented a higher sensitivity to the occurrence of the damage compared with the linear one, justifying the application of a nonlinear metric when the system exhibits intrinsically nonlinear behavior. 
\end{abstract}

\begin{keyword}
\texttt{ Uncertainties \sep damage detection \sep 
	stochastic Volterra model \sep nonlinear behavior.}
\end{keyword}

\end{frontmatter}



\section{Introduction}
\label{Introduction}
The damage detection is the first, and maybe, the most crucial step in the Structural Health Monitoring (SHM) problems \cite{Rytter1993}. So, before trying to locate, quantify and predict the progression of the damage, it is imperative to detect it with some level of confidence. Although a large number of works done in this area of research, some issues have not yet been answered. So, this work focus on two main issues that can complicate the damage detection problem.

The first one is associated with the presence of uncertainties in the measured data. The real structures are subjected to the presence of uncertainties that can be revealed in data change when experimental tests are performed. This data variation can be related to many different factors such as noise in the measurements, a variation of temperature, changes in boundary conditions and sensors/actuators positions, among others \cite{soize2012stochastic,Soize20132379,Soize2017}. Additionally, the damage features and indexes can be sensitive to the data change, confusing the damage detection problem and becoming required the statistical analysis to predict the real condition of the structures with a probabilistic confidence \cite{WORDEN2018139,MAO2013333}. To overcome this issue, several works have shown the use of Multiple Models (MM) to describe the behavior of linear systems with probabilistic modeling, for example, with applications using Auto-Regressive models \cite{VAMVOUDAKISSTEFANOU2018149}, in non–stationary systems \cite{AVENDANOVALENCIA201759,AVENDANOVALENCIA2017326}, and using transmissibility concept \cite{Sakellariou2019}. So, the present work is related to these in the sense of the construction of a probabilistic model able to represent the variability of the system response in the reference condition.

The second one is the inherent nonlinear behavior during the operational life of the structures. Many authors have adopted nonlinear metrics to detect damages that make linear structures to exhibit nonlinear behavior \cite{LIM2014468,Mandal2015,ZENG2015380}. In this situation, detecting the damage becomes a problem of nonlinear behavior detection, that can be solved using various technics \cite{STC:STC215}. However, a more significant problem is faced when the structure presents nonlinear effects still in the healthy condition, making the classical approaches fail to distinguish the variation related to the nonlinear phenomena to the variation caused by the occurrence of a damage \cite{Bornn2010909}. In this case, the models have to be capable of predicting the intrinsically nonlinear effects of the structure, to perform a better classification of the structural condition.  Seeking to solve the problem related to the structures nonlinear behavior, Shiki et al. (2017) \cite{BruceSHMjournal} have considered a method based on the Volterra series to detect damage in an intrinsically nonlinear structure. The model adopted was based on input/output data, and the results achieved showed to be engaging with a better performance of the index that takes into account the nonlinear phenomenon.  However, the authors did not take into account the data variation correlated to a numerous of uncertainties mentioned in this work, given that only the variation caused by the presence of noise in the measurements was taken into account. 

Additionally, Villani et al. (2018) \cite{VILLANI2018} suggested a new method to detect the presence of damage, considering the initial nonlinear behavior of the structures and the data variation generated by the presence of uncertainties, using a stochastic version of the Volterra series. Two procedures were proposed, the first one based on the Volterra kernels coefficients and the second one based on the kernels contributions. The description of the advantages and properties of both approaches were discussed in the referred article. The strategies were applied in a simulated problem to detect the presence of a breathing crack in a beam that exhibits nonlinear behavior, even in the reference state. The results and the performance of the methods have shown to be promising. However, none experimental application was shown. Therefore, this paper employs the experimental application of the approach based on the stochastic Volterra kernels contributions considering the same setup used by Shiki et al. (2017) \cite{BruceSHMjournal}. The aim is to compare the results obtained through the deterministic and stochastic model and to show the benefit of the use of a stochastic model when the data variations exist. The use of kernels coefficients is not examined here because of the linear characteristic of the emulated damage that does not impact the amplitude of the higher order harmonics, as occurred in Villani et al. (2018) \cite{VILLANI2018}. For more information about the motivation of this work and to become close with the bibliography mentioned to the topic, the reader is invited to consult the companion paper Villani et al. (2018) \cite{VILLANI2018}. This paper shows a brief review of the deterministic and stochastic models used and is focused on the experimental results obtained by the application of the methods.

The paper is organized as follows. Section \ref{Problem statement and hypothesis considered} shows the problem faced and the hypothesis considered in the application of the proposed method. Section \ref{Experimental Setup} presents the experimental setup used in the investigations. Section \ref{The use of deterministic Volterra series to detect damage} shows a review of the mathematical model and the results attained with the use of the deterministic approach proposed by Shiki et al. (2017) \cite{BruceSHMjournal}. Section \ref{The use of stochastic Volterra series to detect damage} shows a review of the mathematical model and the results achieved through the application of the stochastic procedure proposed by Villani et al. (2018) \cite{VILLANI2018}. Section \ref{Comparison between the methodologies} brings the comparison between the methodologies used. Finally, section \ref{Final Remarks} presents the discussion about the results and the main conclusion.


\section{Problem statement and hypothesis considered}
\label{Problem statement and hypothesis considered}
This work differentiates two different methodologies to detect damage in initially nonlinear systems, admitting the presence of data variation associated with uncertainties. The first one based on the deterministic Volterra series, expanded using standard Kautz functions, and the second one based on the stochastic version of the Volterra series, expanded applying random Kautz functions. Some attention has to be made to understand correctly this experimental application:
\begin{itemize}
	\item Problem confronted:
	\begin{itemize}
	        \item The system exhibits nonlinear behavior both in reference and damaged conditions;
      		  \item The system response displays random variation generated by the influence of uncertainties;
       		 \item The model has to make difference between uncertainties, nonlinear behavior, and damage.
	\end{itemize}
	\item Experimental random data considered:
		\begin{itemize}
		\item The data variation was not controlled during the experiments. Therefore, it is deemed a random variation;
            \item The system response variation is considered through the restarting of the experimental setup during the different days of tests, to emulate the presence of uncertainties in the measurements. Consequently, the uncertainties are associated to sensors' and actuator's positions, bolts tightening in the clamp, natural temperature fluctuation (this parameter was not controlled), and others related to the assembly and disassembly of the experimental setup;
            \item The measurements were conducted in different days in distinct structural conditions during two weeks;
            \item Additional noise is used in the response samples to generate enough data to be utilized in the Monte Carlo simulations;
            \item The same response samples are analyzed in both methodologies with the goal of comparison.
		\end{itemize}
	\item Input signal used:
		\begin{itemize}
			\item A chirp input is held to identify the model in all conditions because of its nature to excite the structure in a range of frequencies with energy suitable to generate the nonlinear interactions in the response \cite{BruceSHMjournal,REBILLAT20111018,REBILLAT2014247};
            \item The same input is adopted in the identification/training of the models and in the test phase, to assure a reliable identification of the high-order components. The change of the excitation signal nature from the training to the test phase should lead to inferior performance;    
        \end{itemize}    
    \item The damage studied in this work has a linear characteristic (loss of mass). Hence only the method based on the Volterra kernels contributions is analyzed.
\end{itemize}


\section{Experimental Setup}
\label{Experimental Setup}
With the purpose of contrasting the results obtained using different methodologies, the same experimental setup used by Shiki et al. (2017) \cite{BruceSHMjournal} and Villani et al. (2017) \cite{VillaniProcedia2017} is taken into account in this work. The experimental setup consists of: 

\begin{itemize}
	\item A cantilever aluminum beam with 300 $\times$ 18 $\times$ 3 [mm$^3$] of dimension;
	\item A steel mass and a magnet that interacts with each other;
	\item A bolted connection with four nuts;
	\item A MODAL SHOP shaker (Model Number: K2004-E01);
	\item A  vibrometer laser Polytec (Model: OFV-525/-5000-S);
	\item A m+p Vib Pilot data acquisition system.
\end{itemize}

\begin{figure}[!htb]
	\begin{center}
		\subfigure[Experimental apparatus \cite{BruceSHMjournal}.]{\includegraphics[scale=0.6]{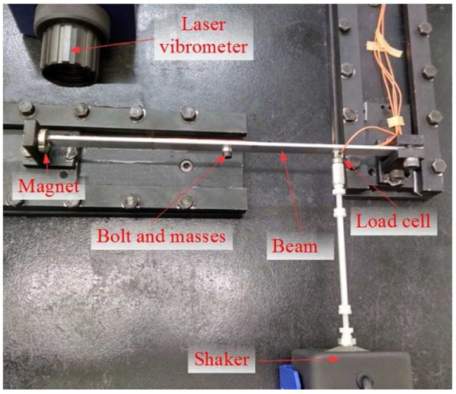}}\quad
		\subfigure[Schematic representation.]{\includegraphics[scale=0.45]{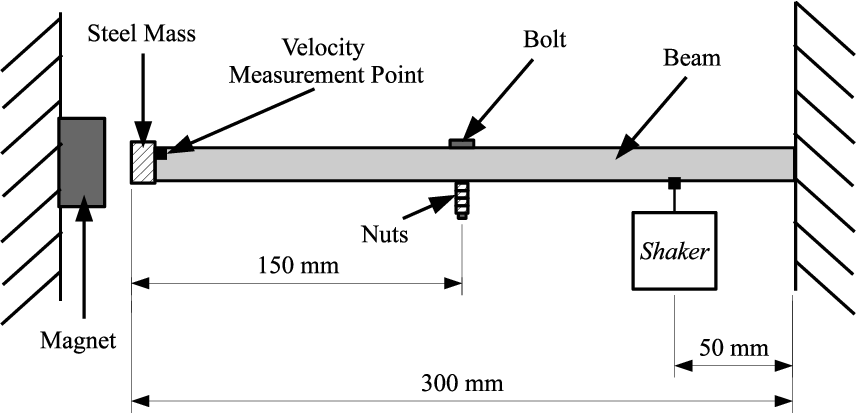}}
		\caption{Illustration of the experimental apparatus used and its schematic representation. Adapted from Shiki et al. (2017) \cite{BruceSHMjournal} and Villani et al. (2017) \cite{VillaniProcedia2017}.}
		\label{mechanical_system_fig}
	\end{center}
\end{figure}

The steel mass is glued in the free extremity of the beam, generating a nonlinear interaction among the beam and the magnet placed near to the free extremity (Fig.~\ref{mechanical_system_fig}). The idea of the experimental setup is emulating a mechanical system with nonlinear behavior due to large displacements. In the context of applications described to damage detection, a bolted connection is allocated $150$ [mm] from the beam free extremity with four nuts with 1 [g] each one. So, the damage inserted is associated with the removal of a mass in bolted connections (nuts loosed). 

The shaker is located 50 mm from the clamp. The structure was excited using different levels of input amplitude 0.01 V (low), 0.10 V (medium) and 0.15 V (high). The vibrometer laser is employed to measure the beam free extremity velocity, that is admitted the system output. The input signal examined in this work is the voltage provided by the amplifier to drive the shaker. It is easier to keep this signal constant over a range of frequency, using the power amplifier in the electrical current mode, than to keep the force applied constant close to the resonance frequency of the system. The same strategy was used in Tang et al. (2016) \cite{Tang2016}. All signals were measured using 1024 Hz as sampling rate, being that 4096 samples were collected. 

\FloatBarrier

\begin{figure}[!htb]
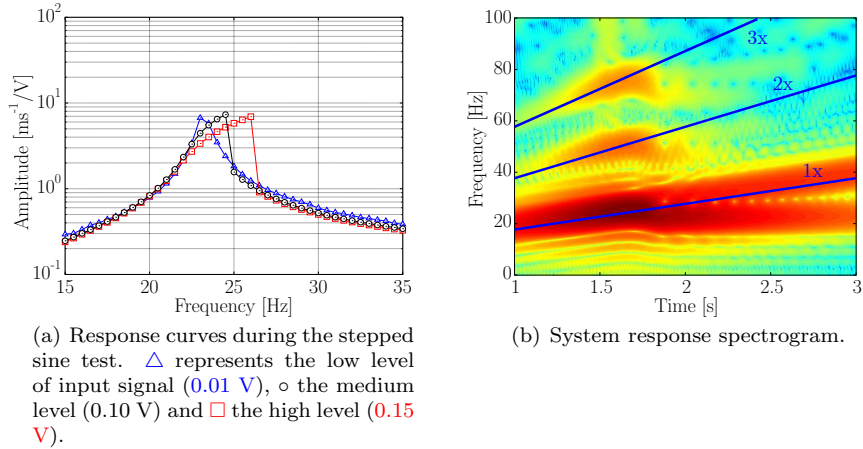

	\fontsize{30}{15}\selectfont
	\begin{center}
		\subfigure[Response curves during the stepped sine test. \textcolor{blue}{{$\triangle$}} represents the low level of input signal 
		(\textcolor{blue}{0.01 V}), \textcolor{black}{$\circ$} the medium level (\textcolor{black}{0.10 V}) and 
		\textcolor{red}{$\square$} the  high level (\textcolor{red}{0.15 V}).]
		{\scalebox{0.25}{\input{Figures/StepAmp.tex}}\label{step_sine_duffing_fig}}\quad
		\subfigure[System response spectrogram.]
		{\scalebox{0.25}{\input{Figures/SpecGram.tex}}\label{SpecGram}}
		\caption{Illustration of the system nonlinear behavior.}
		\label{nonlinear_behavior}
	\end{center}
\end{figure}

The experimental arrangement presents nonlinear behavior with hardening characteristic \cite{kovacic2011duffing}. This characteristic can be seen by the results obtained when the structure was subject to the stepped sine test (Fig.~\ref{step_sine_duffing_fig}). In this test, each sine input signal is applied, and the fundamental amplitude in the stationary state of the output is measured, so, each point of the graphic consider the amplitude of the response obtained from a different time series caused by a sine input signal with a different fundamental frequency. The system response presents the jump phenomenon, related to the nonlinear behavior, that is a result of the large displacements achieved when a high level of input (0.15 V) is applied to the structure. The structure manifests linear behavior when a low level of input (0.01 V) is applied. Additionally, the structure has been excited applying a high level of amplitude (0.15 V) chirp signal, altering the excitation frequency from 10 to 50 Hz (the first mode shape region) in 4 seconds. The spectrogram of the system response can be viewed in Fig.~\ref{SpecGram}.  It is possible to observe the appearance of harmonics of second and third order in the system response. These results attest that the structure exhibits, even in the healthy condition, nonlinear behavior provoked by large displacements.

\begin{figure}[!htb]
	\begin{center}
		\subfigure[Emulated damage \cite{BruceSHMjournal}.]{\includegraphics[scale=1.0]{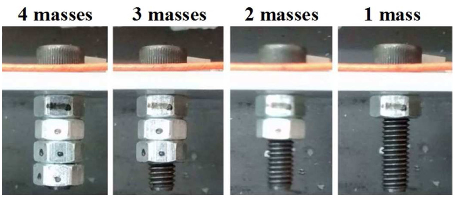}}\\
		\subfigure[Schematic representation of the damage localization.]{\includegraphics[scale=0.65]{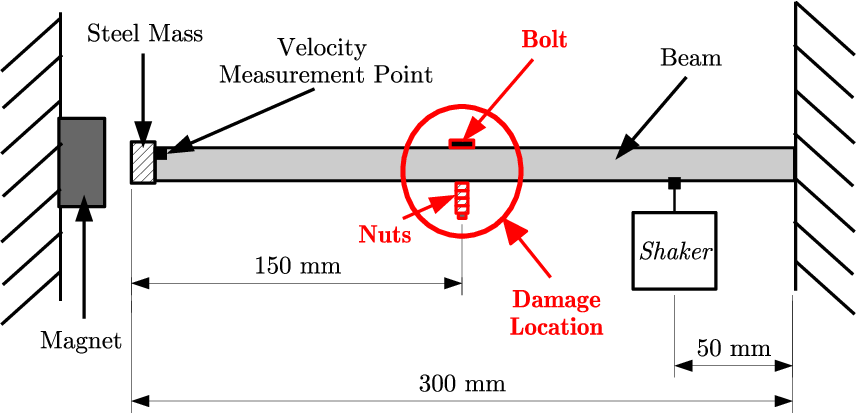}}
		\caption{Illustration of the damage emulated. Adapted from Shiki et al. (2017) \cite{BruceSHMjournal}.}
		\label{mechanical_system_dam_mass}
	\end{center}
\end{figure}

As quoted before, a bolted connection is put in the center of the beam, with four nuts with 0.001 [kg] each one, to emulate the presence of damage, then the healthy state regards the four masses and the damage increases with the loss of the nuts (Fig. \ref{mechanical_system_dam_mass}). Table  \ref{structuralconditions} shows the different structural conditions adopted in this work. Furthermore, the natural variation of the data measured was considered through the repetition of the tests during two weeks in a total of 160 experimental realizations. It was assumed an aleatory variability of the experimental data by restarting the experimental setup and data acquisition in different days, so the variability examined in this work is related to sensors' and actuator's positions, bolts tightening in the clamp, natural temperature fluctuation, and others related to the assembly and disassembly of the experimental setup. Then, Gaussian noise was randomly added to the data, generating a signal to noise ratio (SNR) of 25 dB and a database with 2048 synthetic realizations of the experiments to be held in the stochastic model estimation through the MC simulations \cite{Soize2017,rubinstein2016}. The aim of the immense number of samples measured is to count the data variation related to the measurements made on different days and the noise effect in the data. It is expected that the data fluctuation, caused by the uncertainties such as screws tightening, boundary conditions, sensors position, and others, will not be capable of being described by a deterministic model, confirming the use of a stochastic model to monitor the structural health. This is a real issue that many SHM features have to overcome in practical applications. 

\begin{figure}[!htb]
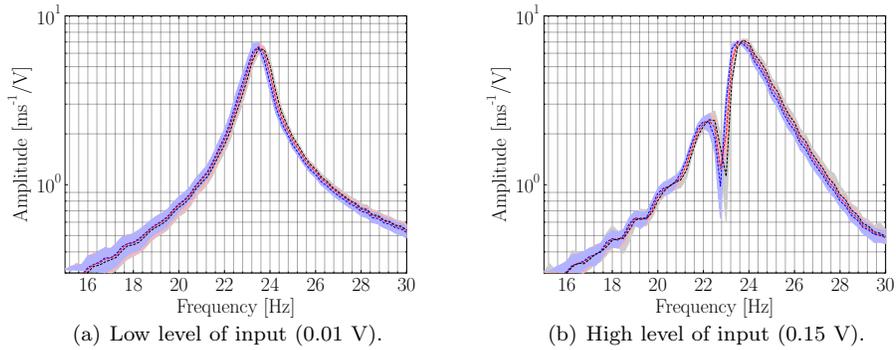

	\fontsize{30}{15}\selectfont
	\begin{center}
		\subfigure[Low level of input (0.01 V).]{\scalebox{0.25}{\input{Figures/FRFsLow.tex}}} \quad
		\subfigure[High level of input (0.15 V).]{\scalebox{0.25}{\input{Figures/FRFsHigh.tex}}} 
				\caption{Frequency Response Function calculated for different structural conditions with 99\% of confidence bands. \colorbox{test}{\textcolor{test}{B}} represents reference condition (4 masses), \colorbox{test2}{\textcolor{test2}{B}} the condition I (3 masses) and \colorbox{gray}{\textcolor{gray}{B}} the condition II (2 masses).}
		\label{FRFsMasses}
	\end{center}
\end{figure}

Figure \ref{FRFsMasses} exemplifies the restriction in detect structural variations (damages) considering the data variation. In this figure, it can be seen the FRFs curves, considering 99\% of statistical confidence bands and different structural states. The clear differentiation between the different conditions is not possible, particularly in the beginning of damage propagation (3 nuts). Therefore, using classical procedures based on deterministic models and methods without any probabilistic analysis of the models or damage indexes, based on single measures, is not reasonable to execute a correct classification of the structural state with probabilistic confidence. Consequently, the method based on the stochastic Volterra series proposed by Villani et al. (2018) \cite{VILLANI2018} can be applied to detect the presence of the damage in these circumstances.

\begin{table}[h]
	\begin{center}
		\caption{Structure conditions.}
		\label{structuralconditions}
		\begin{tabular}{c | c c c c c}
			\hline
			\textbf{State} & H & I & II & III & R \\
			\hline
			\textbf{Condition} &   4 masses (ref.) & 3 masses & 2 masses & 1 mass &  4 masses (repair) \\
			\hline
		\end{tabular}
	\end{center}
\end{table}

\section{Deterministic approach for damage detection}
\label{The use of deterministic Volterra series to detect damage}

This section exposes the application of the damage detection procedure, based on the deterministic Volterra series, proposed by Shiki et al. (2017) \cite{BruceSHMjournal}, regarding the variation on the measured data caused by the appearance of uncertainties. The methodology used is briefly explained, given that the reader can find a more detailed explanation in Shiki et al. (2017) \cite{BruceSHMjournal}. The goal of this section is to prove that the use of a methodology based on a deterministic model, without the use of probabilistic tools and metrics, is not indicated in SHM problems admitting the presence of uncertainties.


\subsection{Deterministic Volterra series}
\label{determ_volt_series}
The deterministic version of the discrete-time Volterra series can be utilized to describe the single output of a causal nonlinear system using the convolution notion \cite{Schetzen}
\begin{equation}
y(k) = \sum_{\eta = 1}^\infty\sum_{n_{1}=0}^{N_{1}-1} \ldots \sum_{n_{\eta}=0}^{N_{\eta}-1} 
\mathcal{H}_{\eta}(n_{1},\ldots,n_{\eta}) \prod_{i=1}^{\eta}u(k-n_{i}),
\label{volt_series_eq}
\end{equation}
where $k \in \Z_+ \mapsto y(k)$ represents the single output that is consequence of the single input $k \in \Z_+ \mapsto u(k)$, $(n_1,..,n_\eta) \in \Z^\eta _+ \mapsto \mathcal{H}_{\eta}(n_{1},\ldots,n_{\eta})$ is the $\eta$-order Volterra kernel and $\Z_+$ represents the set of nonnegative integers. In addition, the greatest advantage of the approach, which will be explored in the present work, is the ability to represent the system output as a sum of the linear and nonlinear components
\begin{equation}
y(k) = \sum_{\eta=1}^{\infty}y_\eta(k) = \underbrace{y_1(k)}_{linear} + \underbrace{y_2(k) + y_3(k) + \cdots}_{nonlinear}.
\end{equation}

However, the use of Volterra series to represent nonlinear system has some shortcomings, being the main the challenge in the series convergence when a large number of terms $N_1,...,N_\eta$ is used. To succeed in this problem, the Volterra kernels $\mathcal{H}_{\eta}$ can be expanded using the Kautz functions \cite{Kautz,heuberger2005modelling}
\begin{eqnarray}
\mathcal{H}_{\eta}(n_{1},...\, ,n_{\eta}) \approx \sum_{i_{1}=1}^{J_{1}}...\sum_{i_{\eta}=1}^{J_{\eta}}
\mathcal{B}_{\eta} \left(i_{1},...\, ,i_{\eta} \right) \prod_{j=1}^{\eta} \psi_{\eta, i_{j}} (n_{j})\,,
\label{kautz_aprox_eq}
\end{eqnarray}
\noindent where $J_{1},\ldots,J_{\eta}$ are the number of Kautz functions used in each orthonormal  projections of the Volterra kernels, $(i_{1},\ldots,i_{\eta}) \in \Z^\eta _+ \mapsto \mathcal{B}_{\eta}(i_{1},\ldots,i_{\eta})$ represents the $\eta$-order Volterra kernel, represented in the Kautz basis, and $n_{j} \in \Z_{+} \mapsto \psi_{\eta, i_{j}} (n_{j})$ represents the $i_{j}$-th Kautz filter. 

So, regarding the Kautz functions approximation (\ref{kautz_aprox_eq}) and the Volterra series model of Eq. (\ref{volt_series_eq}), the system response is reported as
\begin{equation}
y(k)\approx \sum_{\eta = 1}^\infty \sum_{i_{1}=1}^{J_{1}}\ldots\sum_{i_{\eta}=1}^{J_{\eta}}
\mathcal{B}_\eta\left(i_{1},\ldots,i_{\eta}\right) \prod_{j=1}^{\eta} l_{\eta, i_{j}}(k)\, ,
\label{eqresp}
\end{equation}
where $k \in \Z_{+} \mapsto l_{\eta, i_{j}} (k)$ is a simple filtering of the input signal $u(k)$ 
by the Kautz function $\psi_{\eta, i_{j}}$. More information about the nonlinear system identification method based on Volterra series, Kautz functions and the process of Volterra kernels estimation can be found in \cite{daSilva20111103,daSilva2011312,Oliveira2012, BruceSHMjournal,VILLANI2018}. 


\subsection{Damage index}
With the deterministic Volterra model evaluated considering the healthy condition, the system response can be predicted using the approximation (\ref{eqresp}). Then, holding the response predicted by the reference model and the new experimental data measured with the structure in an unknown state, an index can be proposed \cite{BruceSHMjournal}
\begin{equation}
\lambda_{\eta} = \frac{\stddev{e_{\eta,unk}}}{\stddev{e_{\eta,ref}}} \, ,
\end{equation}
where $\lambda_{\eta}$ is the damage index, $\stddev{(.)}$ represents the standard deviation and the prediction errors can be defined as
\begin{eqnarray}
e_{\eta,ref} = y_{exp}^{ref} - \sum_{n = 1}^{\eta} y_n \, ,\\
e_{\eta,unk} = y_{exp}^{unk} - \sum_{n = 1}^{\eta} y_n \, ,
\end{eqnarray}
where $y_{exp}^{ref}$ is the measured response with the structure in the reference condition, $y_{exp}^{unk}$ is the measured response with the system in an unknown state and $y_n$ is the model response contribution of the $n-$order Volterra kernel. In the present work, only the first three Volterra kernels were considered, because of the cubic polynomial characteristic of the nonlinear structure behavior. The statistical analysis of the index has been widely discussed in Shiki et al. (2017) \cite{BruceSHMjournal}, so the reader is encouraged to have an attention to that manuscript for more information.


\subsection{Deterministic model identification}

\begin{figure}[!htb]
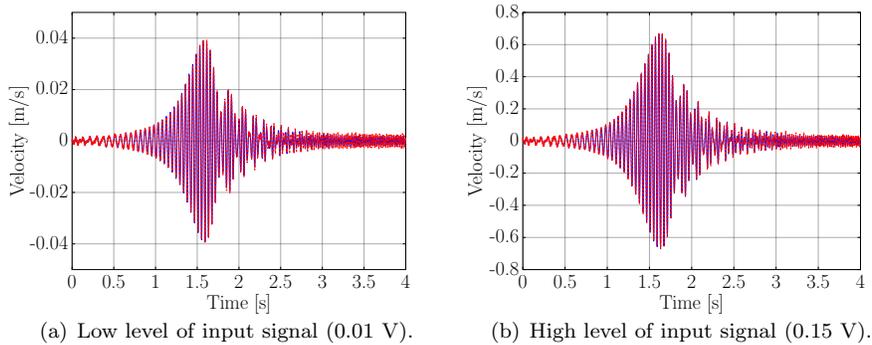

		\fontsize{30}{15}\selectfont
	\begin{center}
		\subfigure[Low level of input signal (0.01 V).]
		{\scalebox{0.25}{\input{Figures/ModeloLowInput.tex}}}\quad
		\subfigure[High level of input signal (0.15 V).]
		{\scalebox{0.25}{\input{Figures/ModeloHighInput.tex}}}
		\caption{Comparison between the system output obtained experimentally and using the Volterra model. The \textcolor{blue}{continuous line ($-$)} represents the output obtained through the Volterra model identified and the \textcolor{red}{dashed line ($--$)} represents the experimental data.}
		\label{ModelResponseDeterministic}
	\end{center}
\end{figure}

The first step to detect damage, considering the proposed method, is to identify a reference model to be applied to the system output prediction. The deterministic Volterra model was identified considering a single sample of response, as the proposed model is deterministic. A chirp signal varying the excitation frequency from $10$ to $50$ Hz was used to excite the structure in the model identification process. As discussed in Shiki et al. (2017) \cite{BruceSHMjournal}, two levels of input amplitude were used to estimate the Volterra kernels in two steps. A low-level input (0.01 V) was used to estimate the first kernel, and then, a high-level input (0.15 V) was employed to evaluate the higher order kernels (second and third). The number of Kautz functions were determined as performed in \cite{BruceSHMjournal}, and defined as $J_1 = 2$, $J_2 = 2$ and $J_3 = 6$. Figure \ref{ModelResponseDeterministic} presents the comparison among the system output obtained experimentally and using the Volterra model identified, considering the two levels of input amplitude and the same chirp signal used in the model identification process. The signals are very similar, revealing that the model can predict the system response in these conditions.

\begin{figure}[!htb]
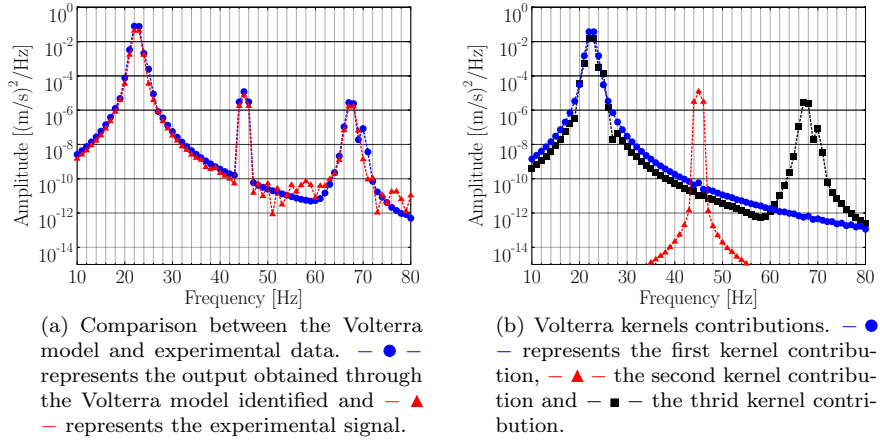

	\fontsize{30}{15}\selectfont
	\begin{center}
		\subfigure[Comparison between the Volterra model and experimental data. \textcolor{blue}{$-$} \mycircle{blue} \textcolor{blue}{$-$} represents the output obtained through the Volterra model identified and \textcolor{red}{$-$} \mytriangle{red} \textcolor{red}{$-$} represents the experimental signal.]
		{\scalebox{0.25}{\input{Figures/ModeloPSD.tex}}}\quad
		\subfigure[Volterra kernels contributions. \textcolor{blue}{$-$} \mycircle{blue} \textcolor{blue}{$-$} represents the first kernel contribution, \textcolor{red}{$-$} \mytriangle{red} \textcolor{red}{$-$} the second kernel contribution and \textbf{$-$} \mysquare{black} \textbf{$-$} the thrid kernel contribution.]
		{\scalebox{0.25}{\input{Figures/ModeloPSDContributions.tex}}}
		\caption{Response obtained considering a sine input with a high amplitude.}
		\label{ModelPSDDeterministic}
	\end{center}
\end{figure}

Then, it is expected to use a different input signal to test the efficiency of the model. So, a single tone sine wave, with a frequency of $23$ Hz that is around the natural frequency of the equivalent linear system, is applied to the structure to test the model performance. Figure \ref{ModelPSDDeterministic}(a) shows the comparison between the model's and the experimental's responses obtained in the frequency domain. So, it can be seen that the model can represent all frequency components. Additionally, Fig. \ref{ModelPSDDeterministic}(b) shows the Volterra kernels contributions. It is remarked that the cubic kernel has a contribution in the first and third harmonics of the response. This point allows the cubic kernel to be susceptible to damages with a linear characteristic. Then, the model can be recognized certified and will be used in the damage detection procedure. 

\subsection{Damage detection}

\begin{figure}[!htb]
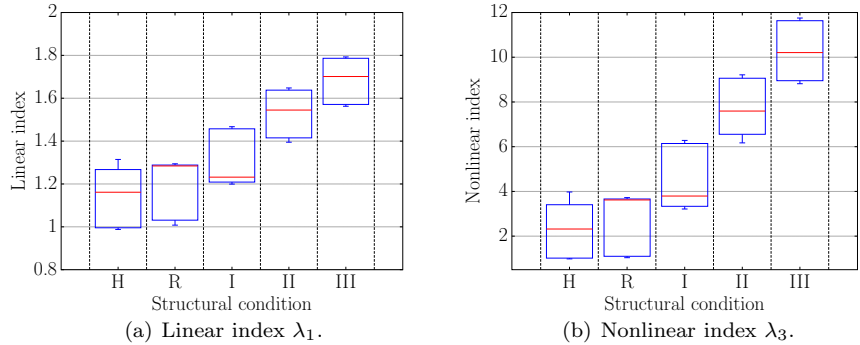

		\fontsize{30}{15}\selectfont
	\begin{center}
		\subfigure[Linear index $\lambda_1$.]
		{\scalebox{0.25}{\input{Figures/IndexLinCoefficients.tex}}}\quad
		\subfigure[Nonlinear index $\lambda_3$.]
		{\scalebox{0.25}{\input{Figures/IndexNLinCoefficients.tex}}}
		\caption{Damage index considering the deterministic model.}
		\label{IndexDeterministic}
	\end{center}
\end{figure}

Now the reference model identified is used in the damage detection procedure. The emulated damage and the conditions considered were described in section \ref{Experimental Setup}. Two damage indexes were calculated, the linear ($\lambda_1$) considering the first kernel and the nonlinear ($\lambda_3$) considering the first three kernels, to compare the results obtained through the linear and the nonlinear approaches. The procedure was applied using a high level of input amplitude, i.e., with the structure operating in a nonlinear regime of motion before the damage occurrence. First of all, Fig. \ref{IndexDeterministic} displays the evolution of the indexes with the increase of the damage. The BoxPlot method is done to conceive more natural the observation of the mean, quartiles and outliers in the indexes distribution, more information about the construction of this graphic can be found in Williams et al. (1989) \cite{BoxPlot}. The nonlinear index presented a higher sensitivity to the damage presence and for both indexes. It is complicated to make difference between the initial propagation of damage (Condition I) and the reference/repair states because it can be observed a superposition in the upper quartile of the indexes computed in the reference state and the lower quartile (even the mean value) of the indexes calculated in the damage I condition.  

\begin{figure}[!htb]
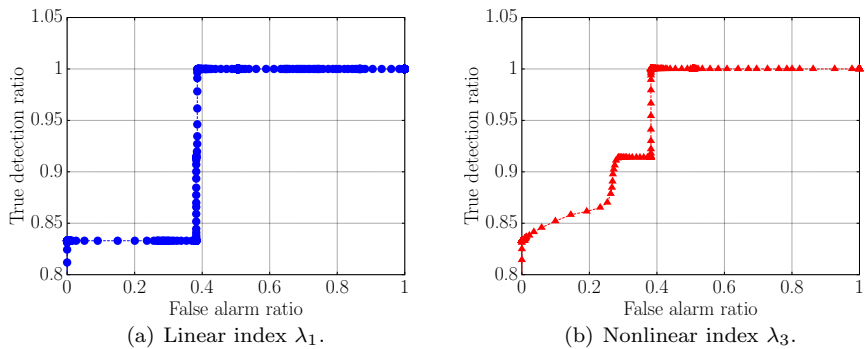

			\fontsize{30}{15}\selectfont
	\begin{center}
		\subfigure[Linear index $\lambda_1$.]
		{\scalebox{0.25}{\input{Figures/ROCLinContribution.tex}}}\quad
		\subfigure[Nonlinear index $\lambda_3$.]
		{\scalebox{0.25}{\input{Figures/ROCNLinContribution.tex}}}
		\caption{ROC curve considering the deterministic model.}
		\label{ROCDeterministic}
	\end{center}
\end{figure}

Finally, to analyze the performance of the indexes to detect damage, the Receiver Operating Characteristics (ROC) curve was computed \cite{farrar2012structural}, considering the two different indexes computed. Figure \ref{ROCDeterministic} shows the results achieved. The nonlinear index has a top performance than the linear one but is evident that both indexes fail to detect the damage in the initial propagation, because of the uncertainties in the data measured in different days. The indexes are not capable of distinguishing the actual data variation, related to the uncertainties, from the presence of the damage. The results shown here are unsatisfactory when we contrast with that one shown in Bruce et. al (2017) \cite{BruceSHMjournal} because the samples used in the present study include more substantial data variation, not only associated with the presence of noise in the measurements but also other variations mentioned before related to uncertainties. Therefore, it is suggested to apply random Volterra series joined with the novelty detection concept. Next section shows the proposed approach and the main results obtained.

\section{Stochastic approach for damage detection}
\label{The use of stochastic Volterra series to detect damage}

The real implementation of SHM approaches, considering the uncertainties, denotes a challenging task, because of the confounding effects which can involve the indexes \cite{MAO2013333}. Some effects like measurement noise, changes in boundary conditions, humidity, temperature, and others, can mask the damages' effects or generate a higher number of false positives \cite{WORDEN2018139}. As a consequence, the measured response, that can vary too much, have to be considered as a random process to ensure that the prediction model can describe the system's behavior with reliability.

So, this section presents the expansion of the deterministic Volterra series theory to a stochastic model and the application on the SHM problem described before, warranting the damage detection even in the presence of data variation and nonlinear behavior. The stochastic model description and methodology proposed is summarized, as the reader can find a detailed mathematical description of the method in the companion paper, Villani et al. (2018) \cite{VILLANI2018}, with a simulated application of the method in full details.


\subsection{The stochastic version of the Volterra series}
\label{The stochastic Volterra series}

A parametric probabilistic approach is employed to describe the system uncertainties in this work, so, the model parameters and processes affected by the uncertainties are considered random parameters and random processes \cite{soize2012stochastic,Soize20132379,Soize2017}. Hence, the parameters and processes are represented on the probability space  $(\SS, \SA, \PM)$, where $\SS$ is sample space, $\SA$ is a $\sigma$-algebra  over $\SS$, and $\PM$ is a probability measure \cite{VILLANI2018}.
Regarding that the system response can fluctuate in the presence of uncertainties, it can be admitted as a random process $(\SSpt,k) \in \SS \times \Z_{+} \mapsto \mathbbl{y}(\SSpt,k)$ illustrated through the random Volterra series as 

\begin{equation}
\mathbbl{y}(\SSpt,k)=\sum_{\eta = 1}^\infty\sum_{n_{1}=0}^{N_{1}-1}\ldots\sum_{n_{\eta}=0}^{N_{\eta}-1}\randvar{H}_{\eta}(\SSpt,n_{1},\ldots,n_{\eta}) \prod_{i=1}^{\eta}u(k-n_{i})\, ,
\label{volt_series_eq_stochastic}
\end{equation}
\noindent where $u(k)$ is a deterministic input signal and $(\SSpt,n_1,..,n_\eta) \in \SS \times \Z^\eta \mapsto \randvar{H}_{\eta}(\SSpt,n_{1},\ldots,n_{\eta})$ 
represents the random version of the $\eta$-order Volterra kernel. 

Besides, the kernels approximation using Kautz functions is recast as

\begin{eqnarray}
\randvar{H}_{\eta}(\SSpt,n_{1},...\, ,n_{\eta})\approx\sum_{i_{1}=1}^{J_{1}}...\sum_{i_{\eta}=1}^{J_{\eta}}\randvar{B}_\eta\left(\SSpt,i_{1},...\, ,i_{\eta}\right)\prod_{j=1}^{\eta}\bbpsi_{\eta, i_{j}}(\SSpt,n_{j})\,,
\label{kautz_aprox_eq_stochastic}
\end{eqnarray}
where $J_{1},\ldots,J_{\eta}$ represents the number of Kautz functions used in the kernels projections, the $\eta$-order random Volterra kernel expanded in the orthonormal basis is represented by the random process
$(\SSpt,i_{1},\ldots,i_{\eta}) \in \SS \times \Z_{+}^\eta \mapsto \randvar{B}_{\eta}(\SSpt,i_{1},\ldots,i_{\eta})$ and $(\SSpt,n_{j}) \in \SS \times \Z_{+} \mapsto \bbpsi_{i_{j}}(\SSpt,n_{j})$ represents the random version of the $i_{j}$-th Kautz filter. As the Kautz functions definition is related to the random system response, they are considered as random processes subjected to the presence of uncertainties.

Then, the approximation (\ref{eqresp}) can be rewritten in a random version

 \begin{equation}
\mathbbl{y}(\SSpt,k)\approx \sum_{\eta = 1}^\infty \sum_{i_{1}=1}^{J_{1}}\ldots\sum_{i_{\eta}=1}^{J_{\eta}}\mathbb B_\eta\left(\SSpt,i_{1},\ldots,i_{\eta}\right)\prod_{j=1}^{\eta} \mathbbl{l}_{\eta, i_{j}}(\SSpt, k)\, ,
\end{equation}
where the random process $(\SSpt,k) \in \SS \times \Z_{+} \mapsto \mathbbl{l}_{i_{j}}(\SSpt, k)$ is a filtering of the deterministic input signal by the random Kautz function. 

Finally, the coefficients of the kernels can be estimated using Monte Carlo simulations and the least squares method. The Monte Carlo method was chosen because it is easier to be implemented when the deterministic algorithm is known. The main drawback is the high number of samples needed to ensure the MC convergence. Then, considering each stochastic execution $\SSpt$, the matrix $\bm{{\Gamma}}$ can be completed with the regressors of the input signal filtered $\mathbbl{l}_{i_{j}}(\SSpt, k)$ and the vector $\mathbf{{y}}$  with the experimental output signal $\mathbbl{y}(\SSpt,k)$

\begin{equation}
{\bm{{\Phi}}}=(\bm{{\Gamma}}^{\scriptsize{\mbox{T}}}\bm{{\Gamma}})^{-1}\bm{{\Gamma}}^{\scriptsize{\mbox{T}}}\textbf{y}\, ,
\label{KernelsEstimation}
\end{equation}
\noindent where ${\bm{\Phi}}$ are the terms of the orthonormal kernels $\mathbb B_\eta$, in each realization $\SSpt$. The procedure is repeated until the Monte Carlo convergence is achieved. More information about the random Kautz functions and the process of the random Volterra kernels estimation can be found in Villani et al. (2018) \cite{VILLANI2018}. 

\subsection{Damage detection based on novelty detection}

On the previous work, two different methods were used, fused with the stochastic Volterra series to detect the presence of a breathing crack in a nonlinear beam, the first one using the random Volterra kernels coefficients and the second one applying the random Volterra kernels contributions. As the simulated damage (a breathing crack model) used in Villani et al. (2018) \cite{VILLANI2018} induced the increase of the second harmonic on the system response, i.e., produced the nonlinear system to present a nonlinear behavior with different nature, the presence of the damage most influenced the second-order Volterra kernels coefficients. However, as stated before,  the use of one or other approach depends on the characteristic of the damage that the structure can be exposed, being the better choice the use of both simultaneously. Additionally, the emulated damage admitted in this article does not influence the amplitude of the harmonic components, only the frequency, causing the kernels coefficients to be unfeeling to the presence of damage. Therefore, the first approach is not adopted here because of this characteristic of the emulated damage, but future works will examine the experimental application using the Volterra series coefficients. As noticed before, only the first three kernels were used to identify the system because of the nonlinear features of the system. 

So, the Volterra kernels contributions in the total system response can be used as damage sensitive index. As cited before, the use of the Volterra series method is supported by its success of separate the nonlinear model response in linear and nonlinear components, considering the kernels identified

\begin{eqnarray}
y_{lin}(k)&\approx& \sum_{i_{1}=1}^{J_{1}} \mathcal{B}_1\left(i_{1}\right) \, l_{i_{1}}(k)\, ,\\
y_{nlin}(k)&\approx& \sum_{i_{1}=1}^{J_{2}}\sum_{i_{2}=1}^{J_{2}} \mathcal{B}_2\left(i_{1},i_{2}\right) \, l_{i_{1}}(k) \, l_{i_{2}}(k) + \,\, \dots \nonumber \\
\dots &+& \sum_{i_{1}=1}^{J_{3}} \sum_{i_{2}=1}^{J_{3}} \sum_{i_{3}=1}^{J_{3}} \mathcal{B}_3\left(i_{1},i_{2},i_{3}\right) \, l_{i_{1}}(k) \, l_{i_{2}}(k) \, l_{i_{3}}(k)\, ,
\end{eqnarray}
\noindent
where $y_{lin}(k)$ and $y_{nlin}(k)$ are the linear and nonlinear contributions, respectively, to the total system response. Then, taken into account the random Volterra series and the MC realizations, the stochastic model is identified in the healthy condition, and the model contributions become a random processes

\begin{eqnarray}
\mathbbl{y}_{lin}(\SSpt,k)&\approx& \sum_{i_{1}=1}^{J_{1}} \randvar{B}_1\left(\SSpt,i_{1}\right) \, \mathbbl{l}_{i_{1}}(\SSpt,k)\, ,\\
\mathbbl{y}_{nlin}(\SSpt,k)&\approx& \sum_{i_{1}=1}^{J_{2}}\sum_{i_{2}=1}^{J_{2}} \randvar{B}_2\left(\SSpt,i_{1},i_{2}\right) \, \mathbbl{l}_{i_{1}}(\SSpt,k) \, \mathbbl{l}_{i_{2}}(\SSpt,k) + \,\, \dots \nonumber \\
\dots &+& \sum_{i_{1}=1}^{J_{3}} \sum_{i_{2}=1}^{J_{3}} \sum_{i_{3}=1}^{J_{3}} \randvar{B}_3\left(\SSpt,i_{1},i_{2},i_{3}\right) \, \mathbbl{l}_{i_{1}}(\SSpt,k) \, \mathbbl{l}_{i_{2}}(\SSpt,k) \, \mathbbl{l}_{i_{3}}(\SSpt,k)\, ,
\end{eqnarray}

\noindent from here, the index $m$ is used to represent the two different indexes calculated. Then, the general notation $\mathbbl{y}_{m}$ is used, with $m = lin$ or $nlin$. As the damage indexes are random processes, the decision about the presence of the damage is executed examining the novelty detection in multivariate data. Taking into account the set of contributions in the healthy condition ($\mathbbl{y}_m$), the standardized Euclidian distance between vectors, can be estimated as

\begin{eqnarray}
\randvar{D}_{m}(\SSpt) = \sum_{n=1}^{N_s} \sqrt{[\mathbbl{y}_{m}(\SSpt,k) - \mathbbl{y}_{m}(n,k)] \, \bm{\Sigma}^{\mbox{-}1} \, [\mathbbl{y}_{m}(\SSpt,k) - \mathbbl{y}_{m}(n,k)]^{\scriptsize{\mbox{T}}} } \, ,
\end{eqnarray}  
\noindent where $\randvar{D}_{m}(\SSpt)$ is the standardized Euclidian distance in reference condition, $\bm{\Sigma}$ is a diagonal matrix in which the diagonal elements are the standard deviations of the columns of $\mathbbl{y}_{m}(\SSpt,k)$. With the aim to concern a threshold to be used in the damage detection procedure, the probability density function (PDF) of the distance calculated in the reference state $\randvar{D}_{m}(\SSpt)$ can be realized using the Kernel Density Estimator \cite{silverman1986density,WORDEN2003323}

\begin{equation}
\hat{p}_{\randvar{D}_{m}}(d_m) = \frac{1}{N_sr} \sum_{i = 1}^{N_s} K \left(\frac{d_{m} - d_{m,i}}{r}\right) \, ,
\label{PDFlambda}
\end{equation}
\noindent  where $\hat{p}_{\randvar{D}_{m}}(d_m)$ is an approximation of the true density $p_{\randvar{D}_{m}}(d_m)$, $d_{m,i}$ is the $i-$th realization of the random variable $\randvar{D}_m(\SSpt)$, $N_s$ represents the number of MC simulations considered, $K$ represents the kernel of the estimator (a Gaussian kernel in the present work) and $r$ is the smoothing parameter.  Then, with the density estimated $\hat{p}_{\randvar{D}_{m}}(d_m)$  it is possible to establish a threshold value \cite{MarcRebillat2016}

\begin{equation}
\mathcal{D}_m = \{d_m \, \,\mbox{such that} \int_{d_m}^{+\infty}  p_{\randvar{D}_{m}}(d_m)\, \mbox{d}d_{m} = \beta\} \, ,
\label{EqThreshold}
\end{equation}

\noindent where $\mathcal{D}_m$ represents the threshold value, and $\beta$ is the sensitivity chosen. Then, to detect damages in the structure, it is proposed to investigate if the set of indexes estimated in the reference condition includes the new contributions obtained through models identified in an unknown structural condition. Therefore, with the structure in an unknown condition (healthy or damaged), a new model has to be identified. The kernels contributions indexes are calculated considering the new model identified ($y_m^{unk}(k)$) and compared with the stochastic reference model, using the same distance-based approach

\begin{eqnarray}
D^{unk}_{m} = \sum_{n=1}^{N_s} \sqrt{[y_m^{unk}(k) - \mathbbl{y}_{m}(n,k)] \, \bm{\Sigma}^{\mbox{-}1} \, [y_m^{unk}(k) - \mathbbl{y}_{m}(n,k)]^{\scriptsize{\mbox{T}}} } \, ,
\end{eqnarray}  

\noindent where $D_{m}^{unk}$ is the standardized Euclidian distance in an unknown condition. Finally, the hypothesis test can be applied to determine if the system is in healthy or damaged condition 

\begin{equation}
\left\{\begin{array}{c}
H_0 :D_{m}^{unk}\leq \mathcal{D}_m\, ,\\
H_1 :D_{m}^{unk}> \mathcal{D}_m\, ,
\end{array}\right.
\label{hypothesis}
\end{equation}

\noindent where the null hypothesis $H_0$ represents the healthy condition and the alternative hypothesis $H_1$ the damaged. The procedure proposed to detect damages can be summarized in 6 main steps, arranged into two phases (training and test) summarized as follows.  
\\

\textbf{TRAINING PHASE:}

\textbf{$-$ Step 1:}  The stochastic Volterra model is identified in the reference condition to construct a set of reference models;

\textbf{$-$ Step 2:} The standardized Euclidian distance ($\randvar{D}_m$) is calculated considering the indexes estimated in the reference condition ($\mathbbl{y}_{m}$);

\textbf{$-$ Step 3:} The threshold value ($\mathcal{D}_m$) is established based on the estimated density of the distances calculated in the reference condition and the probability of false alarms chosen ($\beta$).
\\

\textbf{TEST PHASE:}

\textbf{$-$ Step 4:} A new Volterra model is estimated in an unknown structural condition;

\textbf{$-$ Step 5:} The standardized Euclidian distance (${D}^{unk}_m$) is calculated considering the indexes estimated in an unknown condition (${y}^{unk}_{m}$);

\textbf{$-$ Step 6:} The hypothesis test (Eqs.~\ref{hypothesis}) is applied to compare the distances obtained in the unknown and reference condition.
\\

A more detailed explanation about the approach can be found in Villani et al. (2018) \cite{VILLANI2018}, and in the two next sections, the reader can find the main results obtained through the experimental application of the method to detect damage. 

\subsection{Stochastic model identification}

The Volterra kernels estimation was done in the same form as using the deterministic model, i.e., in two steps considering the same input signal and a low (0.01 V - linear behavior) and a high (0.15 V - nonlinear behavior) level of input. However, in this stochastic application, Monte Carlo simulations were employed to determine the stochastic baseline model, analyzing 2048 realizations to guarantee the method convergence. With the aim of comparing the results of this section with those shown previously, the same number of Volterra kernels and Kautz functions were used (\ref{The use of deterministic Volterra series to detect damage}).

\begin{figure}[!htb]
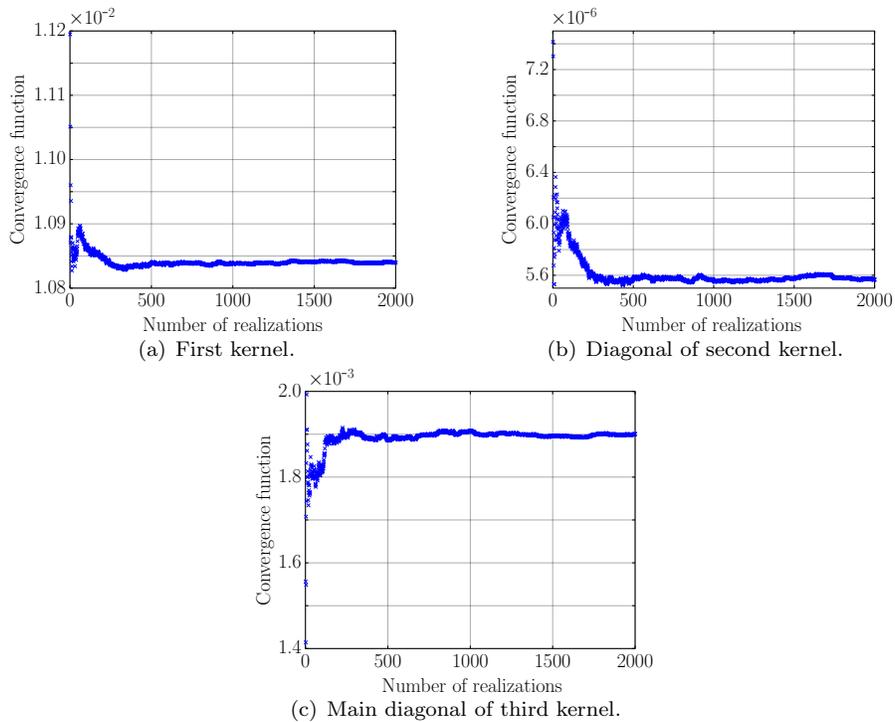

	\fontsize{30}{15}\selectfont
	\begin{center}
		\subfigure[First kernel.]{\scalebox{0.25}{\input{Figures/MCfirstKernel.tex}}} \quad
		\subfigure[Diagonal of second kernel.]{\scalebox{0.25}{\input{Figures/MCsecondKernel.tex}}} \\
		\subfigure[Main diagonal of third kernel.]{\scalebox{0.25}{\input{Figures/MCthirdKernel.tex}}}
		\caption{MC convergence test applied in the Volterra kernels identified in the reference condition (4 nuts).}
		\label{ConvKernelsLinearDamage} 
	\end{center}
\end{figure}

Initially, the convergence of MC simulations has to be studied, ensuring the statistical reliability of the obtained results. The study was done considering a function that depends on the Volterra kernels estimated, defined by

\begin{equation}
conv(N) = \sqrt{\frac{1}{N}\sum_{n=1}^{N}\int_{t=t_0}^{t_f} || {\mathbbl{h}}(\SSpt_n,t) ||^2dt} \, ,
\end{equation}

\noindent where $N$ denotes the number of MC realizations, $\mathbbl{h}(\SSpt_n,t)$ represents the $n$-th realization of the random first kernel or main diagonal of the high order kernels and $||.||$ expresses the standard Euclidean norm. The reader can obtain more details about the method in Soize (2005) \cite{Soize2005623}. Figure \ref{ConvKernelsLinearDamage} shows the results obtained with the criterion applied to the first three kernels identified. It can be noted that the convergence was achieved with 2048 samples used. 

\begin{figure}[!htb]
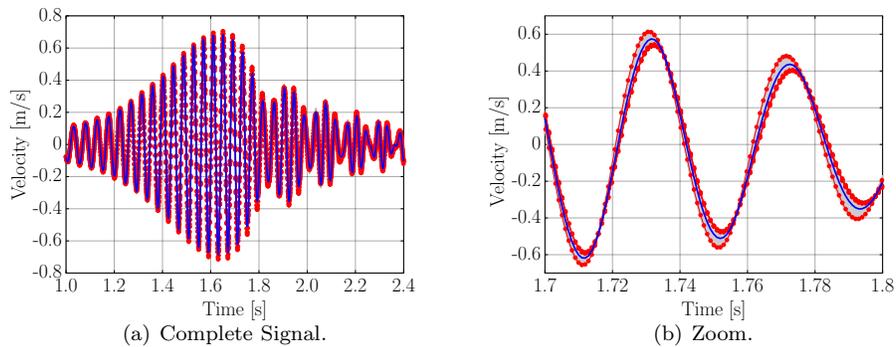

			\fontsize{30}{15}\selectfont
	\begin{center}
		\subfigure[Complete Signal.]{\scalebox{0.25}{\input{Figures/RespHighRandom.tex}}} \quad
		\subfigure[Zoom.]{\scalebox{0.25}{\input{Figures/RespHigh2Random.tex}}} 
		\caption{Comparison between the response obtained through the stochastic Volterra model estimated and obtained experimentally, regarding a high-level of input amplitude  (0.15 V) and the reference condition. The gray box represents 99\% of confidence bands, \textcolor{blue}{-- --} represents the mean and \textcolor{red}{-- $\circ$ --} the experimental data.}
		\label{RespChirpHighLinear}	
\end{center}
\end{figure}

\begin{figure}[!htb]
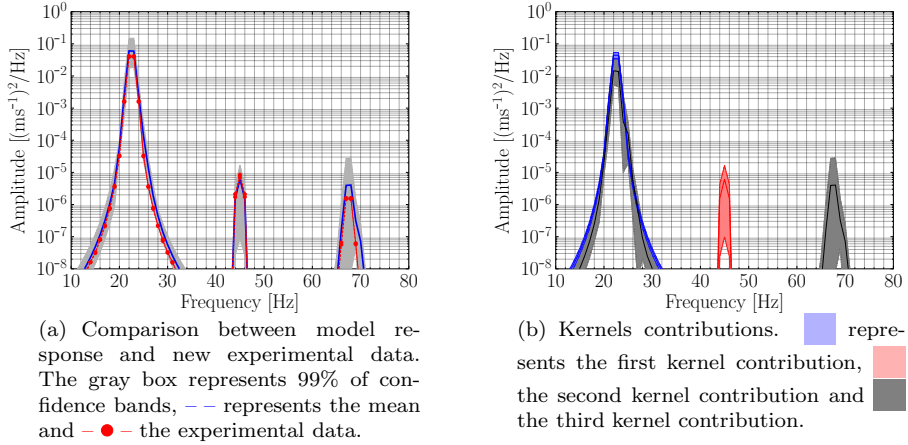

	\fontsize{30}{15}\selectfont
	\begin{center}
		\subfigure[Comparison between model response and new experimental data. The gray box represents 99\% of confidence bands, \textcolor{blue}{-- --} represents the mean and \textcolor{red}{--} \mycircle{red} \textcolor{red}{--} the experimental data.]{\scalebox{0.25}{\input{Figures/PsdHighRandom.tex}}} \quad
		\subfigure[Kernels contributions. \colorbox{test}{\textcolor{test}{B}} represents the first kernel contribution, \colorbox{test2}{\textcolor{test2}{B}} the second kernel contribution and \colorbox{gray}{\textcolor{gray}{B}} the third kernel contribution.]{\scalebox{0.25}{ \input{Figures/PsdHighContributionRandom.tex}}} 
		\caption{Frequency components of the stochastic Volterra model response in comparison with new experimental data, considering a sine input and the reference condition.}
		\label{RespSineLinear}
	\end{center}
\end{figure}

After that, the reference model was verified using two different input signals. First, the same chirp signal examined in the model estimation was applied, and then, a sine wave with a frequency of 23 Hz (first structural fundamental frequency) was adopted. Both signals were applied with a high level of input amplitude (0.15 V) to explore the nonlinear behavior of the system. Figure \ref{RespChirpHighLinear} shows the results obtained using the chirp input with the 99 \% confidence bands. The stochastic model can describe the system behavior with probabilistic confidence. Additionally, Fig. \ref{RespSineLinear} shows the results reached for the sine input, in the frequency domain, to help the analysis of the harmonic components. It is observed that the 99\% model's confidence bands can describe all the system frequency components and the data variation. The kernels contributions reveal that the first kernel has a contribution to the primary harmonic, the quadratic kernel has a contribution to the second harmonic, and the third kernel has a contribution to the principal and third harmonics. This result points that the cubic kernel is sensitive to linear variations, which allows the third kernel to detect damages with linear characteristics, as the loss of mass considered in this work.

\begin{figure}[!htb]
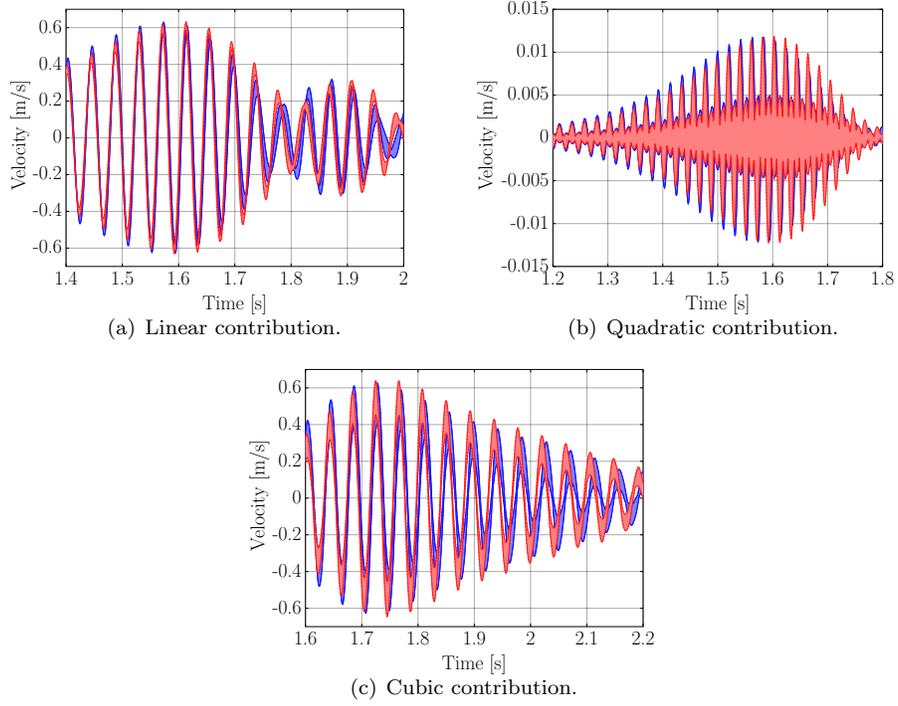

	\fontsize{30}{15}\selectfont
	\begin{center}
		\subfigure[Linear contribution.]{\scalebox{0.25}{\input{Figures/ContributionLinearHigh.tex}}} \quad
		\subfigure[Quadratic contribution.]{\scalebox{0.25}{\input{Figures/ContributionQuadraticHigh.tex}}} \\
		\subfigure[Cubic contribution.]{\scalebox{0.25}{\input{Figures/ContributionCubicHigh.tex}}} 
		\caption{Volterra kernels contributions with 99\% of statistical confidence bands, using a high level of input (0.15 V). \colorbox{test}{\textcolor{test}{B}} represents the condition H and \colorbox{test2}{\textcolor{test2}{B}} the condition III.}
		\label{RespChirpContributionHighLinear}
	\end{center}
\end{figure}

Finally, Fig. \ref{RespChirpContributionHighLinear}  shows the Volterra kernels contributions with 99\% of statistical bands, considering a high level (0.15 V) of the input signal amplitude and two structural conditions, reference (4 masses) and severe damage (1 mass).  The linear and cubic contributions manifest significant differences with the occurrence of the damage, but the second kernel contribution is not so influenced. The variation of the cubic kernel contribution is associated with the influence of the Kautz functions variation that provides varies even with the linear characteristic of the damage. It is also important to recognize that the cubic kernel influences the primary harmonic of the response, so if the presence of the damage alters the first harmonic's behavior, the cubic kernel is changed too. It is clear that in the heavy damage condition, the difference between the kernels contributions in the reference and damaged states can be seen, but when the damage is low and, considering the data variation caused by uncertainties, it is more complicated to detect the damage. So, the indexes were estimated to help the judgment about the structural condition. Next section explains the main results of the application of the damage detection proposition.

\subsection{Damage detection using the stochastic model}

\begin{figure}[!htb]
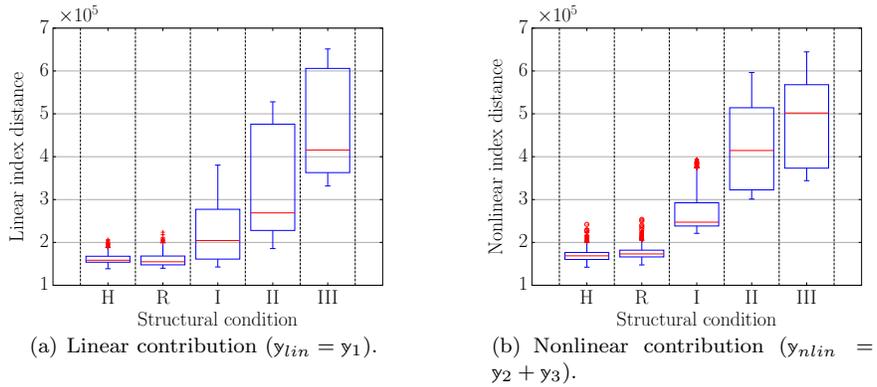

		\fontsize{30}{15}\selectfont
	\begin{center}
		\subfigure[Linear contribution ($\mathbbl{y}_{lin} = \mathbbl{y}_{1}$).]{\scalebox{0.25}{\input{Figures/IndexLinContribution.tex}}} \quad
		\subfigure[Nonlinear contribution ($\mathbbl{y}_{nlin} = \mathbbl{y}_{2} + \mathbbl{y}_{3}$).]{\scalebox{0.25}{\input{Figures/IndexNLinContribution.tex}}} 
		\caption{Damage index (standardized Euclidian distance) considering the stochastic approach.}
		\label{DamageIndexContributionLinear}
	\end{center}
\end{figure}

With the random model identified and verified in the reference condition, it is feasible to apply the novelty detection approach proposed. The excitation signal assumed in this step is the same used in the kernels identification process, using a high-level amplitude (0.15 V), aiming the analysis considering the nonlinear effects in the response. Figure \ref{DamageIndexContributionLinear} shows the growth of the indexes, linear and nonlinear, subjected to the standardized Euclidian distance, with the progression of the damage. The linear index has large dispersion, mainly with the progress of the damage, this behavior illustrates that the presence of uncertainties more influences the linear contribution, so, to detect the damage based on the linear kernel contribution is more difficult, although the damage has a linear characteristic. The nonlinear index is more affected by the effect of the damage, because the third kernel has a contribution in the fundamental frequency of system oscillation, and the presence of uncertainties does not so influence it. Additionally, it can be seen some superposition between the red dots of $\mathbb{D}_{nlin}$ calculated in the reference condition and the lower quartile of the damage I condition. However, these dots represent outliers with no statistical relevance in the performance of the metric as will be seen just ahead.

\begin{figure}[!htb]
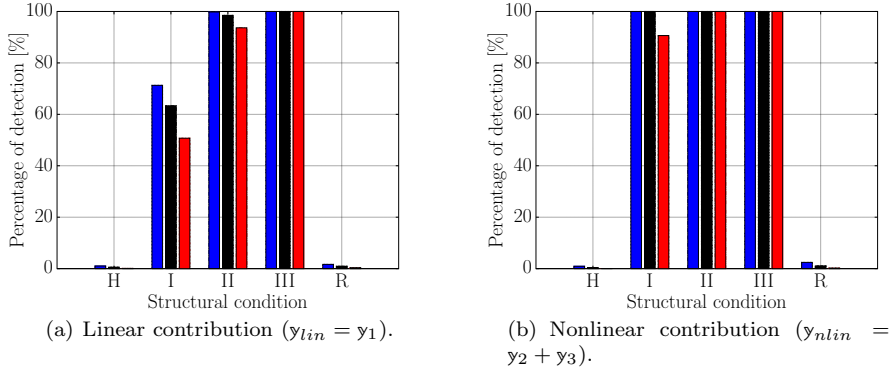

		\fontsize{30}{15}\selectfont
	\begin{center}
		\subfigure[Linear contribution ($\mathbbl{y}_{lin} = \mathbbl{y}_{1}$).]{\scalebox{0.25}{\input{Figures/PODLinContribution.tex}}} \quad
		\subfigure[Nonlinear contribution ($\mathbbl{y}_{nlin} = \mathbbl{y}_{2} + \mathbbl{y}_{3}$).]{\scalebox{0.25}{\input{Figures/PODNLinContribution.tex}}}
		\caption{Percentage of damage detection obtained using the kernels contributions, considering different structural conditions and probability of false alarm used. \colorbox{blue}{\textcolor{blue}{B}} represents $\beta = 0.02$, \colorbox{black}{\textcolor{black}{B}} represents $\beta = 0.01$ and \colorbox{red}{\textcolor{red}{B}} represents $\beta = 0.005$.}
		\label{PODContributionLinear} 
	\end{center}
\end{figure}

It is important to recognize that the estimation of the Volterra kernels was made in two steps, i.e., the linear kernel is estimated considering a low level of response amplitude, and in this condition, the influence of the uncertainties is significant. Consequently, because of this, the linear kernel is more influenced by the confounding effects admitted to the damage detection process. The fact of the cubic kernel be more sensitive to the presence of damage makes the nonlinear index to be capable of differing the variation related to the presence of damage to the variation related to the uncertainties. To exemplify better the capability of the indexes to detect the structural variation related to the damage, the hypothesis test was applied. After the computation of the distances in the reference condition, the threshold value is determined based on the probability of false alarm ($\beta$) chosen, and the distribution achieved.  It was considered three different values for the probability of false alarms ($\beta = 0.005$, $0.01$ and $0.02$) to exemplify the capability of damage detection. The hypothesis test was applied and the results obtained are pointed in Fig. \ref{PODContributionLinear}. As supposed, the performance of the nonlinear index is better. It is pointed out that the linear index was not able to classify right the different conditions evaluated, presenting a lower level of detection.

\begin{figure}[!htb]
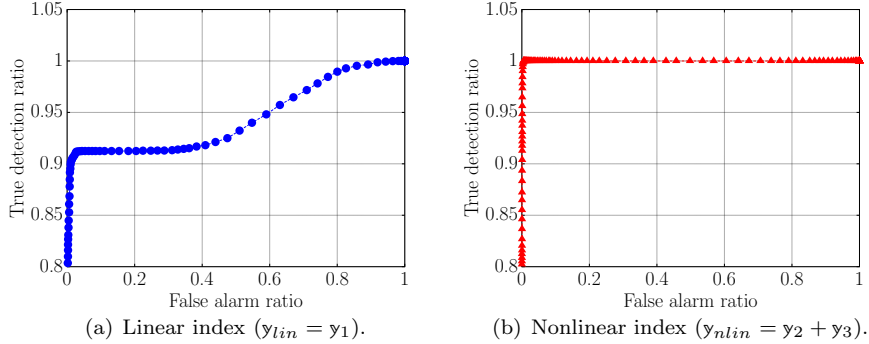

		\fontsize{30}{15}\selectfont
	\begin{center}
		\subfigure[Linear index ($\mathbbl{y}_{lin} = \mathbbl{y}_{1}$).]{\scalebox{0.25}{\input{Figures/ROCLinContributionRandom.tex}}}\quad
		\subfigure[Nonlinear index ($\mathbbl{y}_{nlin} = \mathbbl{y}_{2} + \mathbbl{y}_{3}$).]{\scalebox{0.25}{\input{Figures/ROCNLinContributionRandom.tex}}}
		\caption{ROC curve considering the stochastic model.}
		\label{ROCContributionLinear}
	\end{center}
\end{figure}

Again, the ROC curve was calculated to analyze the performance of the approach better and to make a more direct comparison between the deterministic and stochastic method. Figure \ref{ROCContributionLinear} brings the results for the linear and nonlinear index. The performance of the nonlinear index is better than the linear one, as it is shown with the higher level of precise detection and lower level of false alarms concerned using the nonlinear index. Then, even though the linear feature of the damage imposed (mass variation) the nonlinear index exhibited better results because the nonlinear coefficients are more sensitive to the damage, with a lower impact of the uncertainties. Additionally, when we confront these results with those obtained with the deterministic model, a large improvement in the capability of damage detection considering the nonlinear index can be regarded, with a higher number of true detection and a lower number of false alarms when we examine the nonlinear index. Hence, it can be observed that the use of the stochastic version of the Volterra series outlines an evolution in the field of damage detection regarding the nonlinear behavior and the uncertainties if we compare Fig. \ref{ROCContributionLinear} and Fig. \ref{ROCDeterministic}. 

\section{Comparison between the methodologies}
\label{Comparison between the methodologies}
With the goal of comparison between the methodologies used, Tab. \ref{Comparison} shows the main characteristics of both metrics. As discussed before, the stochastic method proposed improves the capability of the model to detect damage considering the data variation in the response of the initially nonlinear system. This improvement is associated with the ability of the method to "learn" with the data variation in the training phase to predict the response with statistic confidence.

On the other hand, the stochastic methodology requires a higher number of experimental data in the training phase to assure the convergence of the Monte Carlo method and the statistical confidence of the reference model. This characteristic leads to longer processing time. So, the choice of the process to be used it depends on the level of uncertainties/data variation that the system is bared.

\begin{table}[h]
	\begin{center}
		\caption{Main characteristics of both methodologies.}
		\label{Comparison}
		\begin{tabular}{L{3.3cm} | C{3.8cm} R{3.8cm}}
			\hline
			\textbf{Methodology} & \textbf{Deterministic} & \textbf{Stochastic} \\
			\hline
			\textbf{Reference model} &   One single model  & Set of reference models \\
			\hline
			\textbf{Volterra kernels} &   Deterministic -- $\mathcal{H}$  & Random -- $\randvar{H}$ \\
			\hline
			\textbf{Kautz functions} &   Deterministic -- $\psi$  & Random -- $\bbpsi$ \\
			\hline
			\textbf{Model output} &   A single signal -- $y$ & Set of signals predicting the response with statistical confidence -- $\mathbbl{y}$ \\
			\hline
			\textbf{Damage detection} &   The reference model response is compared with the system response in different conditions  & The model is identified in each condition and the contributions are compared \\
			\hline
			\textbf{Damage index} &   Prediction error based method  & Distance-based method \\
			\hline
			\textbf{Experimental data required to estimate the model} &   One single signal measured in reference condition  & A high number of data measured in reference condition to ensure the Monte Carlo convergence  \\
			\hline
			\textbf{Processing time to estimate the reference model} &   Lower  & Higher \\
			\hline
			\textbf{Performance} &   Poor performance considering the data variation  & High performance considering the data variation \\
			\hline
		\end{tabular}
	\end{center}
\end{table}

\section{Summary and Conclusions}
\label{Final Remarks}

This work has compared the use of the deterministic Volterra series methodology and a new stochastic version of the series to detect damage in an initially nonlinear system, regarding the data variation occasioned by the presence of uncertainties. The methodologies were applied in an experimental test using a clamped-free beam operating in a nonlinear regime of motion because of the influence of a magnet positioned near to its free extremity. The damage was emulated through the loss of mass in a bolted connection placed in the center of the beam. Although the damage has a linear characteristic, with influence in the natural frequency of the equivalent linear system, the nonlinear indexes have shown to be more sensitive to detect it when the system is running in a nonlinear regime of vibration, as presumed by previous results \cite{BruceSHMjournal}.  

As the newest result, the use of the stochastic Volterra kernels contributions mixed with the novelty detection metric offered a more significant capability to detect the damage when the data deviation, related to the measurements performed in different days, was admitted. So, this experimental application has shown the effectiveness of the proposed strategy to detect structural variations considering the intrinsically nonlinear effect and change in the data measured. The stochastic Volterra coefficients were not used in this work as damage sensitive feature because of the linear characteristic of the damage examined, as stated before. Though, for future works, the authors aim to apply the stochastic approach, including the use of the kernels coefficients, to detect damage with nonlinear characteristic (a breathing crack) experimentally, considering the structure intrinsically nonlinear behavior and the data variation. 

\section*{Acknowledgments}
The authors are thankful for the financial support provided from of S\~ao Paulo  Research Foundation (FAPESP), grant numbers 2012/09135-3, 2015/25676-2, 2017/24977-4 and 2017/15512-8, from Carlos Chagas Filho Research Foundation of Rio de Janeiro State (FAPERJ) under grants E-26/010.002.178/2015 and E-26/010.000.805/2018, Coordena\c{c}\~ao de Aperfei\c{c}oamento de Pessoal de N\'ivel Superior - Brasil (CAPES) - Finance Code 001,  and CNPq grant number 307520/2016-1. The authors would also like to thank the valuable comments provided by the reviewers and the Associate Editor.


\bibliography{SHMreferences}

\end{document}